\documentclass[preprint,pra,nofootinbib]{revtex4}
\usepackage[mathscr]{euscript}
\usepackage{graphicx}
\usepackage{float}
\usepackage{dcolumn}
\usepackage{bm}
\usepackage{graphicx}
\usepackage{amsmath,amssymb,amsthm}
\usepackage{subfigure}
\usepackage{color}
\usepackage{hyperref}
\hypersetup{colorlinks=true,linkcolor=blue}

\newcommand{\bea}{\begin{eqnarray}}
\newcommand{\eea}{\end{eqnarray}}
\newcommand{\beq}{\begin{equation}}
\newcommand{\eeq}{\end{equation}}

\def\/{\over}

\begin{document}

\title{Optical soliton formation controlled by angle-twisting in photonic moir\'{e} lattices}

\author{Qidong Fu,${}^{1}$ Peng Wang,${}^{1}$  Changming Huang,${}^{2}$ Yaroslav V. Kartashov,${}^{3,4}$ Lluis Torner,${}^{3,5}$ Vladimir V. Konotop,${}^{6,7}$ Fangwei Ye$^{1\ast}$\\
\emph{${}^{1}$School of Physics and Astronomy, Shanghai Jiao Tong University, Shanghai 200240, China}\\
\emph{${}^{2}$Department of Electronic Information and Physics, Changzhi University, Shanxi 046011, China}\\ 	
\emph{${}^{3}$ICFO-Institut de Ciencies Fotoniques, The Barcelona Institute of Science and Technology, 08860 Castelldefels (Barcelona), Spain}\\
\emph{${}^{4}$Institute of Spectroscopy, Russian Academy of Sciences, Troitsk, Moscow, 108840, Russia}\\
\emph{${}^{5}$Universitat Politecnica de Catalunya, 08034 Barcelona, Spain}\\
\emph{${}^{6}$Departamento de F\'{i}sica, Faculdade de Ci\^encias, Universidade de Lisboa, Campo Grande, Ed. C8, Lisboa 1749-016, Portugal}\\
\emph{${}^{7}$Centro de F\'{i}sica Te\'{o}rica e Computacional, Universidade de Lisboa, Campo Grande, Ed. C8, Lisboa 1749-016, Portugal}\\
\emph{$^\ast$Corresponding author: fangweiye@sjtu.edu.cn}
}

\date{\today}


\maketitle

\textbf{Exploration of the impact of synthetic material landscapes featuring tunable geometrical properties on physical processes is a research direction of the highest current interest because of the outstanding phenomena that are continuously uncovered. Twistronics and the properties of wave excitations in moir\'e lattices are salient examples. Moiré patterns bridge the gap between aperiodic structures and perfect crystals, thus opening the door to the exploration of effects accompanying the transition from commensurate to incommensurate phases. Moir\'{e} patterns have revealed profound effects in graphene-based systems~\cite{GrapheneMoire,Woods14,BandFlatGraphene,UnconvenSupercond,Science}, they are used to manipulate ultracold atoms~\cite{atomic1,atomic2} and to create gauge potentials~\cite{Gauge}, and are observed in colloidal clusters~\cite{colloids}. Recently, it was shown that photonic moir\'e lattices enable the observation of the two-dimensional localization-to-delocalization transition of light in purely linear systems~\cite{LDTscirep,LDTnature}. Here we employ moir\'e lattices optically-induced in photorefractive nonlinear media~\cite{Efremidis02,Fleischer03,Freedman06} to elucidate the formation of optical solitons under different geometrical conditions controlled by the twisting angle between the constitutive sublattices. We observe the formation of solitons in lattices that smoothly transit from fully periodic geometries to aperiodic ones, with threshold properties that are a pristine direct manifestation of flat-band physics~\cite{LDTnature}.}

By and large, the linear transport and localization properties of excitations in a material are intimately determined by its inner symmetry and geometrical properties, including a periodic or aperiodic nature, as it occurs in electronic \cite{electron01}, atomic \cite{matter01,matter02}, optical \cite{optical01,optical02}, or two-dimensional material  \cite{graphene01} systems. This feature is directly related to the nature of the eigenstates of the system, which can be extended or localized. When an underlying material exhibits nonlinear response, formation of self-sustained excitation, alias solitons, becomes possible \cite{solitons01,solitons02,solitons03}. Nevertheless, the properties of solitons are still strongly impacted by the linear spectrum of the system. Optical media offer a unique laboratory for the investigation of solitons in different environments. Thus, two-dimensional self-trapping and soliton formation have been investigated in fully periodic optical lattices~\cite{Fleischer03,Yang03,Efremidis03,Neshev03}, as well as in quasicrystals, which are characterized by broken translational invariance~\cite{Freedman06,Ablowitz06,Law10,Ablowitz12,Denz10}. However, formation of solitons at the transition from aperiodic to periodic systems has been never explored experimentally because of lacking of a suitable setting. Optical moir\'{e} lattices~\cite{LDTnature} offer a powerful platform enabling such study. Created with incommensurate geometries, moir\'e patterns may allow localization of light even in the linear limit due to the existence of a large number of extremely flat bands in their spectra, a property that therefore strongly impacts the diffraction of beams in such media. Thus, since soliton states can emerge due to a balance between diffraction and self-phase-modulation induced by nonlinearity, moir\'e lattices allow the investigation of the formation of solitons controlled by the geometry of the induced optical potential. Here we provide the experimental evidence of such possibility, by reporting qualitative differences in soliton excitation dynamics in commensurate and incommensurate moir\'e lattices.

\begin{figure}[ht]
\centering
\includegraphics[width=\linewidth]{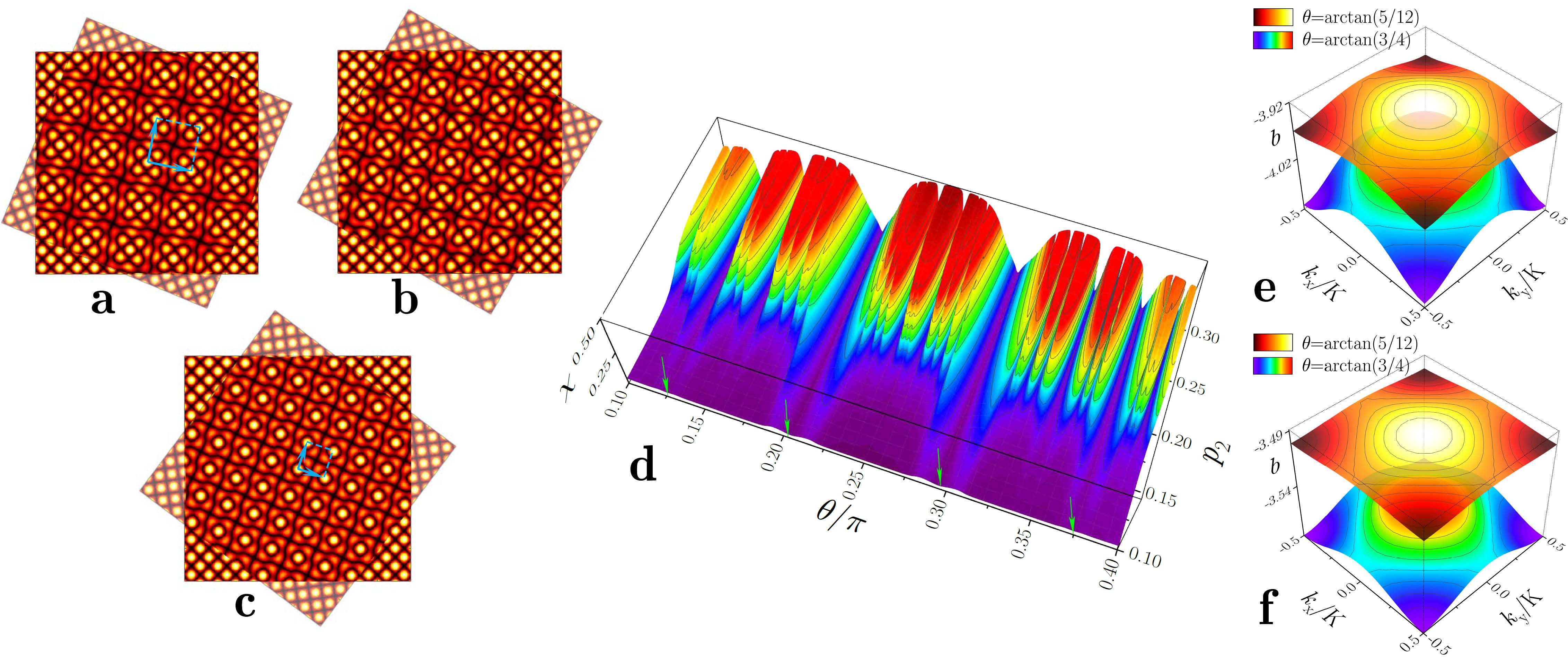}
\caption{
\setlength{\baselineskip}{10pt}
{\bf Moir\'{e} patterns and properties of their linear eigenmodes}. Example of periodic ({\bf a,c}) and aperiodic ({\bf b}) moir\'{e} lattices $I({\bm r}_\perp)$ produced by two superimposed square sublattices with $p_1=0.5$ and $p_2=0.3$ rotated by the angles $\theta=\textrm{arctan}(5/12)$ ({\bf a}), $\theta \approx 0.167\pi$ ({\bf b}), $\theta=\textrm{arctan}(3/4)$ ({\bf c}). Blue arrows in ({\bf a}),({\bf c}) indicate the primitive lattice vectors. ({\bf d}) Form-factor (inverse width) of the most localized linear eigenmode of the lattice versus rotation angle $\theta$ and depth $p_2$ of the second sublattice at $p_1=0.5$. Green arrows indicate angles corresponding to the (3,4,5) and (5,12,13) Pythagorean triples. The top lattice bands for two Pythagorean angles $\theta=\arctan(3/4)$ and $\theta=\arctan(5/12)$ superimposed in one plot for $p_2=0.1$ ({\bf e}) and $p_2=0.3$ ({\bf f}). Bloch momenta $k_{x,y}$ are normalized to the width of the Brillouin zone equal to $K \approx 1.265$ for $\theta=\arctan(3/4)$ and to $K \approx 0.785$ for $\theta=\arctan(5/12)$.}
\label{Figure1}
\end{figure}

A photonic moir\'{e} lattice is created in a photorefractive crystal by a shallow modulation of the refractive index in the $(x,y)$ plane induced by two mutually rotated, or twisted, periodic square sublattices generated by light with ordinary polarization. The latter is chosen so that the corresponding beams practically do not experience self-action in the crystal, thus propagating undistorted. In contrast, light with extraordinary polarization experiences a strong nonlinear response. Propagation along the $z$-direction of a signal beam in such polarization is governed by the nonlinear Schr\"{o}dinger equation for the dimensionless amplitude $\Psi({\bm r}_{\perp}, z)$~\cite{Efremidis02,Fleischer03}:
\begin{equation}\label{eq1}
\centering
i \frac{\partial \Psi}{\partial z}=-\frac{1}{2}\nabla_{\perp}^2\Psi+\frac{V_0}{1+I(\bm r_{\perp})+ |\Psi| ^2}\Psi.  
\end{equation}
Here  $\nabla_{\perp}=(\partial/\partial x,\partial/\partial y) $; ${\bf r_{\perp}}=(x,y)$ is the radius-vector in the transverse plane; $z$ is the longitudinal coordinate scaled to the characteristic length $2 \pi n_e \lambda$, where $\lambda$ is the wavelength (in our experiments $\lambda=632.8$ nm); $n_e$ is the unperturbed refractive index of the crystal experienced by the extraordinary-polarized light; and $V_0>0$ is the dimensionless applied dc field. Here we set $V_0=5$, which  corresponds to $5.7\times 10^4 $~V/m dc electric field applied to the crystal; $ I({\bm r}_{\perp})=\left|p_1V({\bm r}_{\perp})+p_2V(S{\bm r}_{\perp})\right|^2$ is the moir\'{e} pattern composed by two ordinary-polarized periodic sublattices $V({\bm r}_{\perp})$ and $V(S {\bm r}_{\perp})$ interfering in the $(x, y)$ plane [here $S=S(\theta)$ is the matrix of rotation in the $(x,y)$ plane by the angle $\theta$]; $p_1$ and $p_2$ are the amplitudes of the first and second square sublattices, respectively. Each square sublattice $V({\bm r}_{\perp})$ is formed by the interference of four plane waves~\cite{LDTnature}. In the following we set the amplitude of the first sublattice to $p_1= 0.5$, which  corresponds to an average intensity $I_{\text{av}} \approx 1.9~ \text{mW}/\text{cm}^2$, and tune the amplitude $p_2$ of the second sublattice. For such parameters, the actual refractive index modulation depth in the moir\'{e} pattern illustrated in Fig.~\ref{Figure1}(a-c) is of the order of $\delta n \sim 10^{-4}$.

Two-dimensional Pythagorean moir\'{e} lattices composed of two square Bravais sublattices $p_1 V({\bm r}_{\perp})$ and $p_2 V(S {\bm r}_{\perp})$ (the point group $D_4$) rotated with respect to each other around common lattice site, are periodic (commensurate) structures only when the rotation angle $\theta$ satisfies $\cos\theta = a/c$, $\sin\theta = b/c$, where the positive integers $(a, b, c)$ having no common divisors except $1$, constitute a primitive Pythagorean triple, i.e., $a^2+b^2=c^2$ (Fig.~\ref{Figure1}\textbf{a},\textbf{c}). We call such angles Pythagorean. For other rotation angles the pattern is aperiodic (incommensurate, or almost periodic in mathematical terms), as illustrated in Fig.~\ref{Figure1}\textbf{b}. The linear spectrum of the lattices ~\cite{LDTscirep,LDTnature} can be obtained by omitting the nonlinear term $|\Psi|^2$ in Eq.~(\ref{eq1}) and searching for the corresponding linear eigenmodes in the form $\Psi({\bm r}_{\perp}, z)=\psi({\bm r}_{\perp})e^{i b z}$, where $b$ is the linear propagation constant and $\psi({\bm r}_{\perp})$ is the transverse field distribution. To characterize the mode localization we use the integral form-factor $\chi =(\iint |\Psi|^4 d^2 {\bm r}_{\perp})^{1/2}/U$, with $U$ being the mode power $U=\iint |\Psi|^2 d^2 {\bm r}_{\perp}$. The larger $\chi$ the stronger the localization. 

The dependence of the  form-factor of the mode with the largest $b$ (this is the most localized mode) on $\theta$ and $p_2$ is shown in Fig.~\ref{Figure1}\textbf{d}. For Pythagorean angles the mode is delocalized for any depth $p_2$ of the second sublattice because in this case the moir\'{e} pattern is periodic, but for non-Pythagorean angles the mode becomes localized if $p_2$ exceeds some critical value, $p_2^{\rm cr} \approx 0.18$ corresponding to the linear localization-delocalization threshold. The physical origin of this phenomenon is the suppressed diffraction due to flatness of the allowed bands of the effective Pythagorean lattice approximating real incommensurate moir\'{e} pattern at $p_2>p_2^{\rm cr}$ \cite{LDTnature}. This observation comes from a general rule: the higher the order of the primitive Pythagorean triple (determined by the integer $c$) the larger the area of the respective primitive cell of the lattice (see the blue arrows in Fig.~\ref{Figure1}\textbf{a},\textbf{c}) and the smaller the width (in $b$) of the allowed bands, thus  indicating a reduced diffraction strength. This behavior is illustrated in Fig.~\ref{Figure1}\textbf{e},\textbf{f} that compares top bands of the Floquet-Bloch spectra calculated for two different Pythagorean angles and two different sublattice amplitudes $p_2$. As discussed below, the angular-dependent band flattening  has direct implications for soliton formation in the lattices, because it exposes that the diffraction strength experienced by narrow linear inputs in the lattices notably decreases with increase of the order of the primitive Pythagorean triple. This is particularly clear at $p_2>p_2^{\rm cr}$, where patterns akin to discrete diffraction are observed in the linear limit and is less pronounced at $p_2<p_2^{\rm cr}$, when such patterns co-exist with a rapidly expanding broader background.

\begin{figure}[h]
\includegraphics[width=0.7\linewidth]{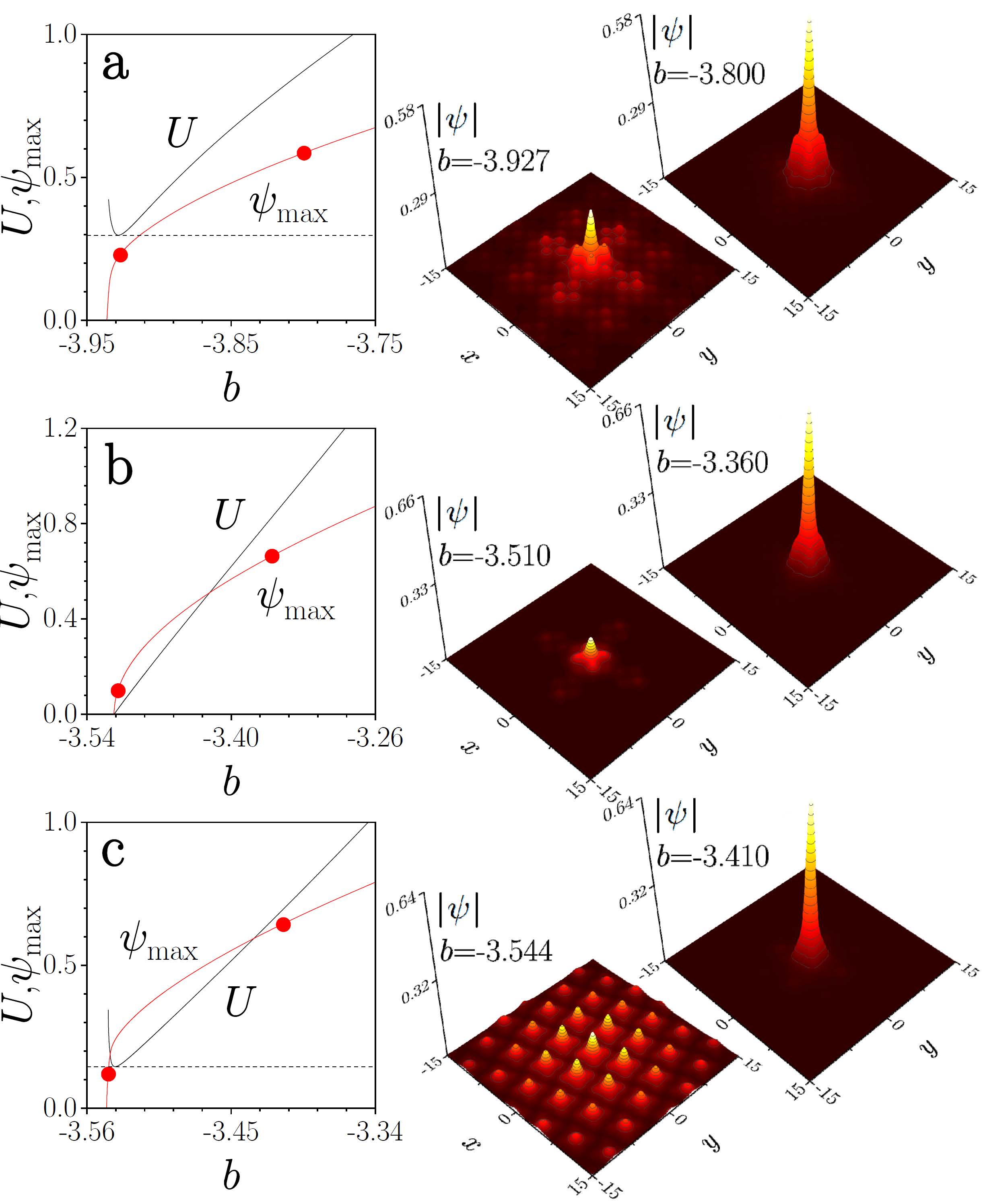}
\caption{
\setlength{\baselineskip}{10pt}
{\bf Families of two-dimensional solitons in moir\'{e} lattices}. Soliton power $U$ and peak amplitude $\psi_\text{max}$ versus propagation constant $b$ (left column \textbf{a}, \textbf{b}, \textbf{c}) and representative soliton profiles (right column) are shown for incommensurate  moir\'{e} lattice with the rotation angle $\theta \approx 0.139\pi$ below ($p_2=0.1$, the upper panels) and above ($p_2=0.3$, the middle panels) the critical value $p_2^{\rm cr}$, as well as for a lattice with the Pythagorean rotation  angle  $\theta=\arctan(3/4)$ with  $p_2=0.3$  (bottom  panels). The profiles shown in the right column correspond to red dots on the $\psi_\text{max}(b)$ curves.
}
\label{Figure2}
\end{figure}

Turning now to the nonlinear regime, we look for soliton solutions of Eq.~(\ref{eq1}) in the form $\Psi({\bm r}_{\perp}, z)=\psi({\bm r}_{\perp}) e^{ibz}$, where $b$ is the nonlinear propagation constant, which for $V_0>0$ (focusing nonlinearity) exceeds the propagation constants of the linear eigenmodes. Solitons form families characterized by the dependencies of the peak amplitude $\psi_{\text{max}}=\text{max} {|\psi|}$ and power $U$ on $b$, as shown in Fig.~\ref{Figure2}. Since incommensurate moir\'e lattices have either delocalized (at $p_2 < p_2^{\rm cr}$) or localized (at $p_2 > p_2^{\rm cr}$) linear modes, solitons in such lattices show a completely different behaviour in the low-amplitude limit depending on the $p_2$ value. When linear localized modes do not exist ($p_2 < p_2^{\rm cr}$), then consistent with the behavior of homogeneous media~\cite{Townes}, solitons in incommensurate moir\'e lattices can exist only if they carry a power $U$ that exceeds a certain threshold value $U_{\text{th}}$ (Fig.~\ref{Figure2}\textbf{a}) below which they quickly diffract. However, when $p_2>p_2^{\rm cr}$ and linear localized states exist, the soliton family bifurcates from the respective linear localized mode, remaining well-localized at any power $U$. Under such conditions, solitons do not feature a power existence threshold (Fig.~\ref{Figure2}\textbf{b}). In contrast, in commensurate moir\'{e} lattices corresponding to the Pythagorean angles, solitons always exhibit a nonzero existence power threshold for any depth of the second sublattice $p_2$ (Fig.~\ref{Figure2}\textbf{c}), and strongly expand at low amplitudes. In all the cases shown in Fig.~\ref{Figure2}, the peak soliton amplitude vanishes in the linear limit (see red curves in Fig.~\ref{Figure2}\textbf{a}-\textbf{c}), while far from it the solitons become strongly localized. All shown soliton families are either completely stable, when power threshold is zero, or unstable at low amplitudes and get stabilized at high amplitudes, when the power threshold does not vanish.

\begin{figure}[h]
\centering
\includegraphics[width=0.7\linewidth]{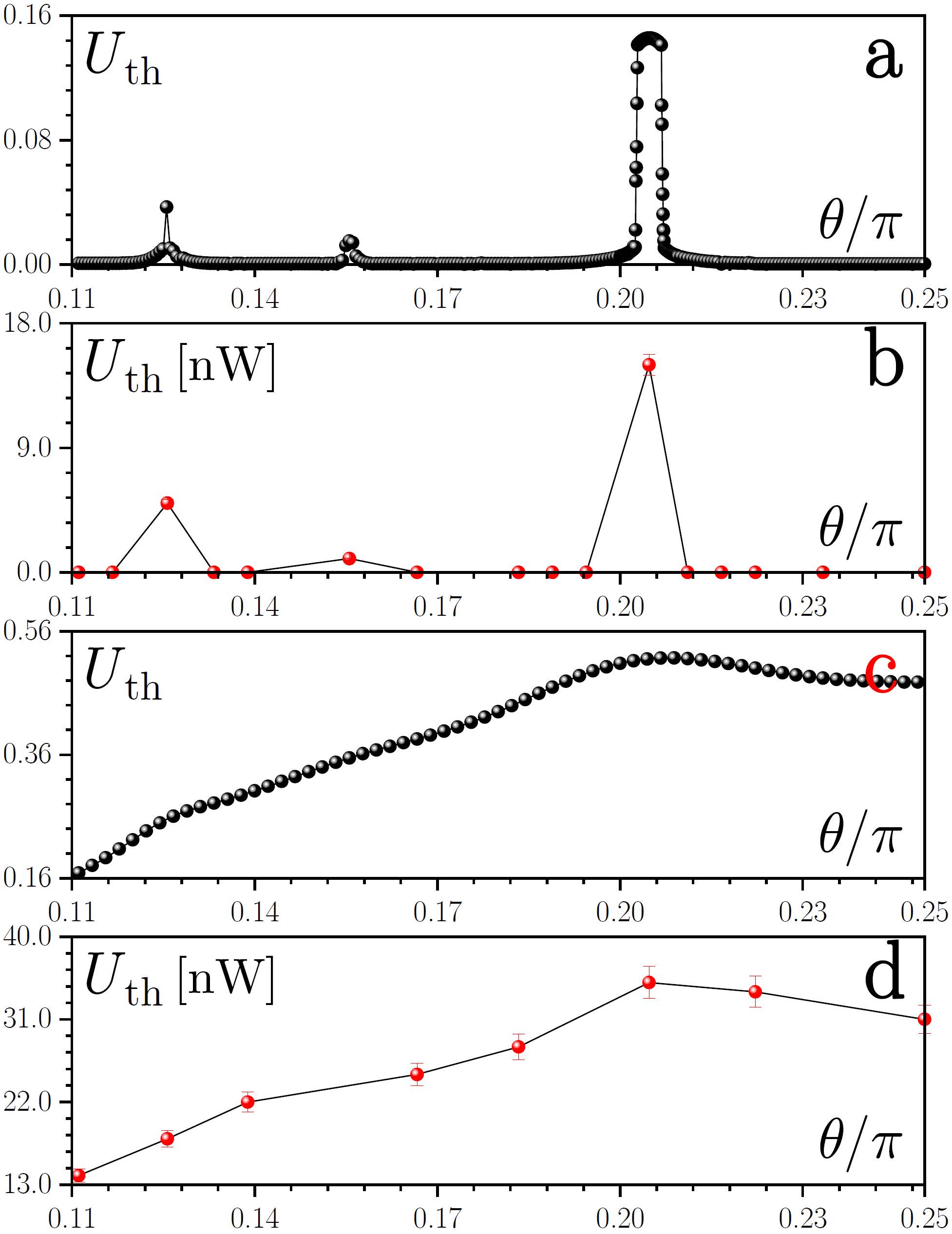}
\caption{
\setlength{\baselineskip}{10pt}
{\bf Thresholds for soliton formation in moir\'{e} lattices}. Comparison of theoretically calculated ({\bf a},{\bf c}) and experimentally measured ({\bf b},{\bf d}) dependencies of threshold for soliton formation on rotation angle $\theta$ for the depth of the second sublattice above localization-delocalization threshold $p_2=0.3$ ({\bf a},{\bf b}) and below localization-delocalization threshold $p_2=0.1$ ({\bf c},{\bf d}).
}
\label{Figure3}
\end{figure}

Here we report the experimental observation of all such scenarios using lattices with readily tunable twisting angle. More specifically, the transition between commensurate and incommensurate moir\'{e} lattices or between the regimes, where delocalization and localization takes place in the linear limit, can be explored by adjusting the phase mask used for the creation of the sublattices or by tuning their relative amplitudes. The comparison between the theoretical predictions and the experimental observations of the threshold power $U_{\text{th}}$ as a function of the rotation angle  $\theta$ between the sublattices is shown in Fig.~\ref{Figure3}. For $p_2$ above the critical value (Fig.~\ref{Figure3}\textbf{a}), $U_{\text{th}}$ vanishes for non-Pythagorean angles while it exhibits narrow peaks around the Pythagorean angles. As visible in Fig.~\ref{Figure1}\textbf{e},\textbf{f}, the band curvature decreases with the increase of the order of the Pythagorean triple. Thus, the nonlinearity required to balance the curvature-induced diffraction decreases too, in agreement with the observations depicted in  Fig.~\ref{Figure3}\textbf{a},\textbf{b}. The first, second, and third highest peaks occur near the Pythagorean angles corresponding to the primitive triples $(3, 4, 5)$, $(5, 12, 13)$, and $(8, 15, 17)$, respectively, in agreement with the properties of the bands at such angles. Also, the heights of the peaks are observed to diminish as areas of the primitive cells of the respective moir\'e patterns increases. Notice that the area of a Pythagorean cell, that is defined by $c$, non-monotonically depends on the twist angle. The smallest area of the primitive (square) cell of a Pythagorean moir\'e pattern corresponds to the primitive triple $(3, 4, 5)$, that explains the locations of the absolute maxima in all panels of Fig.~\ref{Figure3}. Because the bands become flatter when the areas of the primitive cells increase, other $U_\textrm{th}$ maxima are almost undetectable. This is consistent with the approximation of incommensurate Pythagorean moir\'e patterns by commensurate ones established in~\cite{LDTnature}. In all experimental results reported, as well as in all numerical simulations showing dynamical soliton excitation, we observed stability of two-dimensional beams. 

Remarkably, the above relation between the order of the Pythagorean triple associated with the lattice and the threshold for soliton formation, even though less pronounced, was still observed in the regime $p_2 < p_2^{\rm cr}$, where localization in the linear system is impossible even in incommensurate lattices (Fig.~\ref{Figure3}\textbf{c},\textbf{d}). Under these conditions the soliton formation threshold $U_\textrm{th}$ does not vanish for any rotation angles, although it remains sensitive to the diffraction properties of the lattice. Our observations show that it achieves the maximal value around $\theta=\arctan(3/4)$ angle that corresponds to the commensurate Pythagorean lattice associated with the $(3,4,5)$ triple and having the smallest possible primitive cell. In all cases we observed an excellent agreement between the theoretical predictions (Fig.~\ref{Figure3}\textbf{a},\textbf{c}) and the experimental (Fig.~\ref{Figure3}\textbf{b},\textbf{d}) results.

The threshold power $U_{\text{th}}$ for soliton formation was measured by evaluating the soliton content  $C=U_\text{out}/U_\text{in}$, where $U_\text{in}$ is the power of the input beam measured at $z=0$ and $U_\text{out}$ is the power of the beam that remains at the output after $2$ cm of propagation, within a pinhole of radius of $27~ \mu$m approximately equal to one period of the sublattice forming the moir\'e pattern. In the numerical modeling of the dynamical soliton excitation within the frames of the model (\ref{eq1}), we used the same criterion calculating the output power for Gaussian inputs $\Psi=A\, \textrm{exp}(-{\bm r}_{\perp}^2/r_0^2)$ within a  circle of radius equal to the sublattice period, albeit at large propagation distances ($z=1000$) to avoid the presence of transient effects.

\begin{figure}[h]
\centering
\includegraphics[width=\linewidth]{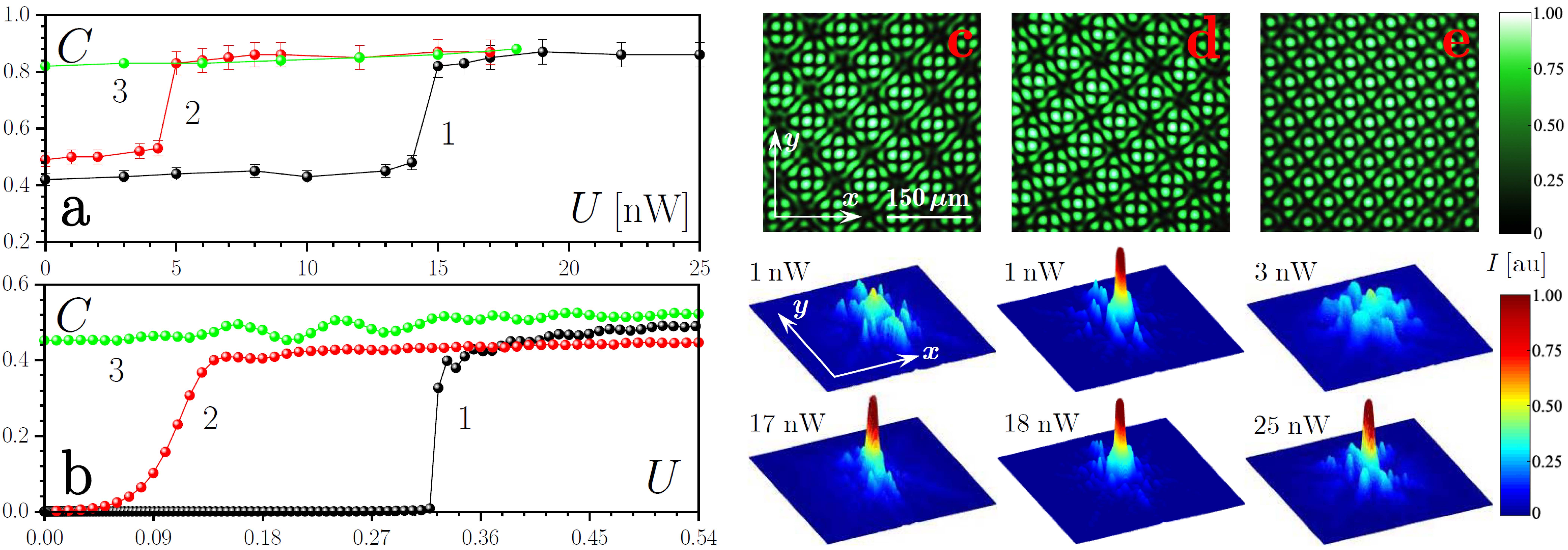}
\caption{
\setlength{\baselineskip}{10pt}
{\bf Soliton formation above the linear localization-delocalization threshold}. Experimentally measured ({\bf a}) and numerical ({\bf b}) soliton content versus input power for Pythagorean rotation angles $\theta=\arctan(3/4)$ (curve 1), $\arctan(5/12)$ (curve 2), and for non-Pythagorean one $0.139\pi$ (curve 3). The second sublattice depth is $p_2$=0.3. Lattice profiles and corresponding low- and high-power output distributions for experimentally measured signal beam for  ({\bf c}) $\theta=\arctan(5/12)$, ({\bf e}) $\theta=\arctan(3/4)$, and ({\bf d}) $\theta \approx0.139\pi$ angles. Power levels are indicated on the plots.}
\label{Figure4}
\end{figure}

In Fig.~\ref{Figure4} we show the dependence of the soliton content $C$ on the input power $U$, together with the low-power and high-power output beams, for a moir\'e lattice with $p_2 > p_2^{\rm cr}$ (i.e., above the linear localization-delocalization threshold), for different rotation angles. In lattices corresponding to the Pythagorean angles $\arctan(3/4)$ and $\arctan(5/12)$ (black and red lines in Fig.~\ref{Figure4}\textbf{a},\textbf{b}) one observes a sharp jump from diffraction to a soliton content close to unity upon increase of the input power. This means that practically all input power goes into a soliton. The power corresponding to the abrupt increase in $C$ is the dynamical threshold for soliton formation $U_\text{th}$ - it is this quantity that is plotted in Fig.~\ref{Figure3}\textbf{b} and \textbf{d}. For non-Pythagorean angles, i.e. for incommensurate moir\'e lattices, soliton content $C$ remains very high irrespective of the input power (green lines in Fig.~\ref{Figure4}\textbf{a},\textbf{b}) because such lattice supports linear modes that are effectively excited by the input beam. Figures~\ref{Figure4}\textbf{c}-\textbf{e} illustrate that for Pythagorean angles the output beam becomes localized only at sufficiently high power, while for non-Pythagorean angles the output is always localized, consistent with theoretical predictions and with expectations based on the flatness of the bands at such twisting angles. 

\begin{figure}[h]
\centering
\includegraphics[width=\linewidth]{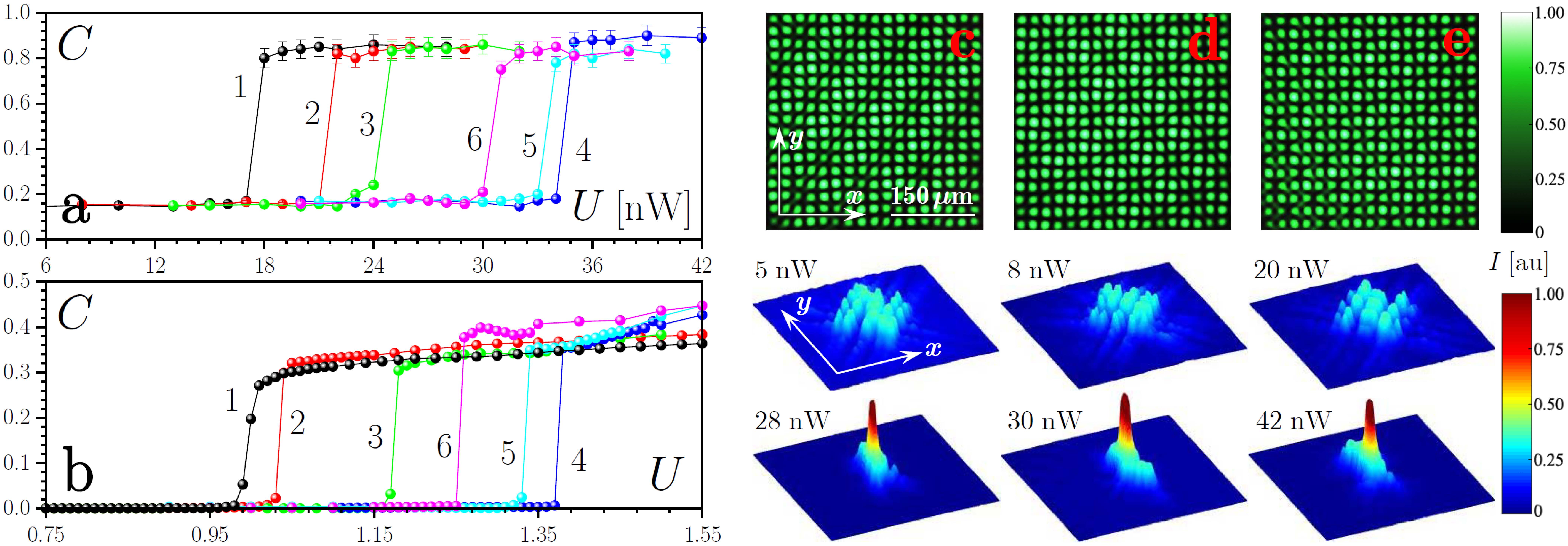}
 \caption{
\setlength{\baselineskip}{10pt}
{\bf Soliton formation below linear localization-delocalization threshold}. Experimentally measured ({\bf a}) and theoretically calculated ({\bf b}) soliton content versus input power for rotation angles $\theta$=arctan(5/12) (curve 1), 0.139$\pi$ (curve 2), 0.167$\pi$ (curve 3), arctan(3/4) (curve 4), 0.222$\pi$ (curve 5), and 0.250$\pi$ (curve 6) for second sublattice depth $p_2$=0.1. Lattice profiles and corresponding low- and high-power output distributions for signal beam for Pythagorean ({\bf c}) $\theta$=arctan(5/12), ({\bf e}) $\theta$=arctan(3/4), and non-Pythagorean ({\bf d}) $\theta \approx0.139\pi$ angles.
}
\label{Figure5}
\end{figure}

Figure \ref{Figure5} depicts the observed behavior of the soliton content for lattices with $p_2 < p_2^{\rm cr}$ that thus cannot support linear localized states. Now we observed step-like behavior of the soliton content $C$ for all Pythagorean and non-Pythagorean angles. Namely, solitons are excited only above a threshold power (see outputs in Fig.~\ref{Figure5}\textbf{a}-\textbf{e}). Notice the higher values of the threshold power relative to the previous cases, consistent with Fig.~\ref{Figure3}\textbf{c} (in all the cases the used input powers are well below the saturation limit). Above threshold, the values of $C$ are similar for different rotation angles, indicating that the excited states are strongly localized on the scale of about one sublattice period.
 
Summarizing, we have observed the excitation of two-dimensional solitons in Pythagorean moir\'e patterns and showed that their excitation dynamics is dictated by the fundamental transition from commensurate to incommensurate lattice geometries. Incommensurability and the relation between the depths of the two sublattices forming the moir\'e pattern has been shown to be crucial for soliton excitation. In all cases, the behavior of the soliton formation threshold is a direct manifestation of the band structure of the moir\'e lattices resulting from the different twisting angles of the sublattices and, in particular, of the general band-flattening associated to the geometry of the moir\'e lattices. We anticipate that similar phenomena may occur in moir\'e patterns composed of sublattices of other crystallographic symmetries, such as twisted honeycomb lattices, and in other physical systems where flat-bands induced by geometry arise.

\section*{METHODS}
 \paragraph*{\bf Experimental setup.}  The experiment is carried out in a biased SBN: 61 photorefractive crystal whose dimensions are 5$\times$5$\times$20 mm$^3$. The experimental setup is similar to the one used in ~\cite{LDTnature}. A continuous-wave laser, with wavelength $\lambda$=532 nm and ordinary polarization, is used to write the desirable moir\'{e} lattices into the sample by superposing two square lattice patterns (sublattices) that have tunable amplitudes $p_{1,2}$ and twist angles $\theta$. To probe the induced moir\'{e} lattice, an extraordinarily polarized beam from He-Ne laser ($\lambda$=632.8 nm) is launched into the sample, and its propagation through the sample is monitored. The power of the probe beam can be varied in the range from several nano-Watts (nW) to micro-Watts ($\mu$W) by using a variable attenuator to control nonlinear self-action due to photorefractive nonlinearity. The beam power in pre-set pinholes with desired radius $R$, connected to the soliton content, is acquired using laser beam profiler.

\paragraph*{\bf Data Availability}
The data that support the findings of this study are available from the corresponding author F. Y. upon reasonable request.
 
\paragraph*{\bf Acknowledgments}
 Q. F., P. W and F. Y acknowledge the support of the National Natural Science Foundation of China under Grant Nos. No. 61475101 and No. 11690033. Y.V.K. and L.T. acknowledge support from the Severo Ochoa Excellence Programme, Fundacio Privada Cellex, Fundacio Privada Mir-Puig, and CERCA/Generalitat de Catalunya. V. V. K. acknowledges financial support from the Portuguese Foundation for Science and Technology (FCT) under Contract no. UIDB/00618/2020.

\paragraph*{\bf Author contributions} 
Q. F and P. W contribute equally to this work. All authors contribute significantly to the work.

\paragraph*{\bf Competing interests} 
The authors declare no competing interests.

\end{document}